
\magnification 1200
\tenpoint
\superrefsfalse
\baselineskip = 0.333 truein \relax
\def\frac#1#2{{#1\over#2}}
\section {Introduction}

The pinch technique (PT) is an algorithm that leads to a
rearrangement of the Feynman graphs contributing to
a gauge-invariant amplitude
(an S-matrix element, for example) in such a way
as to define
individually gauge-independent propagator,vertex, and
box-like structures
\reference{Kursunoglu}
J.~M.~Cornwall,
in Deeper Pathways in High Energy Physics, edited by B.~Kursunoglu,
A.~Perlmutter, and L.~Scott (Plenum, New York, 1977), p.683
\endreference
{}.
Its original motivation
was to devise a consistent truncation scheme for the
Schwinger-Dyson equations
that govern the dynamics of gauge theories. In particular,
it has been extensively employed as a part
of a non-perturbative approach to continuum QCD
\reference{Bib}
J.~M.~Cornwall,
\journal Phys. Rev.; D 26,1453 (1982)
\endreference
\reference{*Hou}
J.~M.~Cornwall, W.~S.~Hou, and J.~E.~King,
\journal Phys. Lett.; B 153,173 (1988)
\endreference
\reference{*Nadkarni}
S.~Nadkarni,
\journal Phys. Rev. Lett; 61,396 (1988)
\endreference
\reference{*C and P}
J.~M.~Cornwall and J.~Papavassiliou,
\journal Phys. Rev.; D 40,3474 (1989)
\endreference
\reference{*Lavelle}
M.~Lavelle,
\journal Phys. Rev.; D 44,26 (1991)
\endreference
\reference{*4gluon}
J.~Papavassiliou,
\journal Phys. Rev.; D 47,4728 (1993)
\endreference
. On the other hand,
most of its recent applications have been motivated by issues
of gauge invariance in a number of one-loop
electro-weak calculations. For instance,
the PT was extended to the case of non-Abelian gauge theories
with Higgs mechanism
in the context of a simplified $SU(2)$ model,
and a gauge independent electromagnetic form factor for the
neutrino was constructed
\reference{Klako}
J.~Papavassiliou,
\journal Phys. Rev.; D 41,3179 (1990)
\endreference
{}.
The application of the PT to the
electro-weak sector of the
Standard Model (SM) was accomplished
in
Ref.\reference{D&S}
G.~Degrassi and A.~Sirlin,
\journal Phys. Rev.; D 46,3104 (1992)
\endreference
. In that paper,
explicit expressions for the one loop
gauge-invariant $WW$, $ZZ$,$\gamma\gamma$, and $\gamma Z$
 self-energies
were derived
in the framework of the renormalizable $R_{\xi}$ gauges,
and an alternative
description of the PT, in terms of
equal time commutators of currents
in relevant Ward identities, was introduced.
The PT was used to compute the anomalous terms induced
in the $WW\gamma$ and  $WWZ$ vertices by one-loop corrections
within the SM
 \reference {Kostas}
J.~Papavassiliou and K.~Philippides,
\journal Phys. Rev.; D 48,4255 (1993)
\endreference
. The
gauge independent
$\gamma W^{+}W^{-}$ and  $ZW^{+}W^{-}$
vertices constructed in this manner
are related to the gauge-independent
 W-self-energies introduced in Ref.\cite {D&S}
by a very simple QED-like Ward identity, and
give rise to magnetic dipole and electric quadrupole form factors,
which, unlike previous
treatments, are gauge independent, infrared finite, and respect
perturbative unitarity.
In Ref.\reference{STU}
G.~Degrassi, B.~Kniehl, and A.~Sirlin,
\journal Phys. Rev.; D 48,R3963 (1993)
\endreference
the SM bosonic contributions to the S,T, and U parameters
\reference{Peskin}
M.~E.~Peskin and T.~Takeuchi,
\journal Phys. Rev. Lett; 65,964 (1990)
\endreference
 were shown to be gauge dependent and,
except when formulated within a restricted class of gauges,
ultraviolet divergent. Within the class of $R_{\xi}$ gauges,
a gauge invariant formulation of the S,T, and U parameters
in the PT framework was proposed, which circumvents the aforementioned
pathologies.
Finally, gauge invariant
magnetic dipole moment form factors for the top-quark
were computed
\reference{Claudio}
J.~Papavassiliou and C.~Parrinello,
NYU/04/06
\endreference
{}.

Given the relative simplicity and apparent
effectiveness of the PT,
questions about its meaning,
the range of its applicability, and the uniqeness of the results obtained,
arise naturally.
As so far the PT has been applied in the framework of the
renormalizable $R_{\xi}$ gauges at the one-loop level, the issue of
whether it has a more general validity is of particular interest.
The aim of this paper is to apply the PT to the derivation of the
one-loop W boson self-energy in the context of the unitary gauge.

We recall that
when the SM is quantized in the framework of the linear
renormalizable $R_{\xi}$ gauges, application of the PT gives
rise to $\xi$-independent
one-loop self-energies for the gauge bosons.
In particular, the PT W self-energy leads to an $SU(2)$
running coupling whose asymptotic behaviour,
as $|q^{2}|\rightarrow \infty$, coincides with that derived from the
renormalization group
\reference{Be'g}
M.~A.~Beg and A.~Sirlin,
\journal Phys. Rep.; 88,1 (1982)
\endreference
{}.
 On the other hand,
since the early days of spontaneously broken non-Abelian gauge theories,
the unitary gauge has been known to give rise to
renormalizable S-matrix elements, but to
Green's functions
that are non-renormalizable in the sense that their divergent parts cannot
be removed by the usual mass and field-renormalization counter-terms
\reference{Weinberg}
S.~Weinberg,
\journal Phys. Rev. Lett; 27,1688 (1972)
\endreference
\reference{SLee}
S.~Y.~Lee,
\journal Phys. Rev.; D 6,1701 (1972)
\endreference
\reference{A&Q}
T.~W.~Appelquist and H.~R.~Quinn,
\journal Phys. Lett; 39B,229 (1972)
\endreference
\reference{Fujikawa}
K.~Fujikawa, B.~W.~Lee, and A.~I.~Sanda
\journal Phys. Rev.; D 6,2923 (1972)
\endreference
{}.
Even though this
situation may be considered acceptable from the
physical point of view,
the inability to define
renormalizable Green's functions has
always been a theoretical shortcoming of the unitary gauge.
In this paper we show that the application of the PT to the unitary
gauge calculations
systematically reorganizes the one-loop S-matrix contributions into
kinematically distinct  pieces (propagators,vertices,boxes) that can
be renormalized with the usual counter-terms characteristic of a
renormalizable theory.
The aforementioned shortcomings
associated with the unitary gauge
are thus circumvented. Furthermore,
the renormalizable amplitudes obtained in this fashion
are {\sl identical} to those calculated in Ref.\cite{D&S}
in the context of the $R_{\xi}$ gauges.

It should be emphasized that the above results,
although welcome and perhaps expected, are by no means obvious.
The point is that the unitary gauge can be obtained from the
$R_{\xi}$ gauges if the limit $\xi\rightarrow\infty$ is taken before
Feynman integrals are performed. Thus, there is no obvious guarantee
that when the PT is applied directly to the highly divergent amplitudes
characteristic of the unitary gauge calculations, it will lead to the same
$\xi$-independent self-energies, vertices, and boxes derived in the
$R_{\xi}$ framework.

The paper is organized as follows: In Section 2 we briefly review the
S-matrix PT and illustrate the method with two useful applications.
In Section 3 we focus on the divergent
contributions in the unitary gauge and show how the PT leads
to a renormalizable
W self-energy. In section 4 we present a detailed proof that the
PT W self-energies (finite as well as divergent parts)
 calculated in the $R_{\xi}$ and unitary gauges, are
identical.

\section {The Pinch Technique.}

The simplest example that illustrates how the P.T. works is the
gluon
two point function.
Consider the $S$-matrix
element $T$ for the elastic scattering of two fermions of masses
$M_{1}$ and $M_{2}$, mediated by gluon exchange.
It is of the form
$\sum_{ij}T^{(ij)}({\bar{u'}}_{1}\Gamma^{(i)} u_{1})
({\bar{u'}}_{2}\Gamma^{(j)} u_{2})$
where the u's are the particle spinors, the $\Gamma^{(i)}$ are
linearly independent operators in spinor space, and the $T^{(ij)}$
are Lorentz invariant cofactors.
 To any order in perturbation theory the  $T^{(ij)}$
are independent
of the gauge fixing parameter $\xi$, defined by the tree-level
gluon propagator
$$
{\cal{D}}_{\mu\nu}(q) =  \frac{-i}{q^2}
[g_{\mu\nu}-(1-\xi)\frac{q_{\mu}q_{\nu}}{q^2}] .
\EQN GaugeProp$$
On the other hand, as an explicit calculation shows,
the conventionally defined proper self-energy (collectively
depicted in graph 1a)
depends on $\xi$. At the one loop level this dependence is canceled by
contributions from other graphs, such as 1b and 1c, which,
at first glance, do not seem to be
propagator-like.
That this cancellation must occur and can be employed to define a
gauge-invariant self-energy, is evident from the decomposition:
$$
T(s,t,M_{1},M_{2})= T_{1}(t) + T_{2}(t,M_{1},M_{2})
+T_{3}(s,t,M_{1},M_{2}) ,
\EQN S-matrix$$
where henceforth we supress the $i,j$ indices. We note that
$T_{1}(t)$ depends only on the Mandelstam variable
$t=-({p'}_{1} -p_{1})^{2}=-q^2$,
 and not on $s=(p_{1}+p_{2})^{2}$ or the
external masses. Typically, self-energy, vertex, and box diagrams
contribute to $T_{1}$, $T_{2}$, and $T_{3}$, respectively.
Moreover, such contributions are $\xi$ dependent. However, as the sum
$T(s,t,M_{1},M_{2})$ is gauge-invariant, it is easy to show that
\Eq{S-matrix} can be recast in the form
$$
T(s,t,M_{1},M_{2})=
{\hat{T}}_{1}(t) + {\hat{T}}_{2}(t,M_{1},M_{2})+
{\hat{T}}_{3}(s,t,M_{1},M_{2}) ,
\EQN S2-matrix$$
where the ${\hat{T}}_{i}$ ($i=1,2,3$) are {\sl separetaly} $\xi$-independent.
  The propagator-like parts of graphs,
which in the PT ensure the gauge independence of ${\hat{T}}_{1}(t)$,
 are called  "pinch parts".
Two examples are given in Fig.(1e) and Fig.(1f), which depict the
pinch parts of diagrams (1b) and (1c), respectively.
They arise whenever a current on an external fermion line is
contracted with its four-momentum, which in turn emerges from
a gluon propagator or an elementary three-gluon vertex.
Such contributions trigger an elementary Ward identity of the form
$$\eqalign{
k^{\mu}\gamma_{\mu} \equiv & \slashchar{k} = (\slashchar{p}+
\slashchar{k}-m)-(\slashchar{p}-m)\cr
=& i[S^{-1}(p+k)-S^{-1}(p)]~.\cr}
\EQN BasicPinch$$
The first term on the r.h.s.
 of \Eq{BasicPinch} removes the
internal fermion propagator - that is a "pinch" - whereas $S^{-1}(p)$
vanishes
on shell. This mechanism is characteristic of the S-matrix P.T. we will use
throughout this paper.
As explained in Ref.\cite{Bib}, it leads to a well-defined prescription
to implement the rearrangement indicated in \Eq{ S2-matrix}.
The gauge-independent function $\hat{T}_{1}(t)$ can then be identified
with the contribution of the new PT propagator.
As emphasized in Ref.\cite{D&S}, in the case of the SM the corresponding
self-energies are endowed with very desirable properties.
Once it has been verified that the PT leads to $\xi$-independent
functions ${\hat{T}}_{1}$, ${\hat{T}}_{2}$, and ${\hat{T}}_{3}$,
any value for the gauge parameter $\xi$ may be chosen in order to determine
their explicit expressions.
Among the $R_{\xi}$ gauges, the Feynman-t'Hooft gauge ($\xi = 1$) is
particularly simple, as it removes the
longitudinal part of the gluon propagator. Therefore, in that gauge
the only possibility for pinching in four-fermion amplitudes
arises from the
four-momenta in the trilinear boson vertices.
Thus, in the case of the gluon-mediated amplitude, the
relevant pinch parts come from Fig.(1b) and its mirror image.
In evaluating these graphs, it is convenient to
split the vertex
$$
\Gamma_{\mu\nu\alpha}(q,k,-k-q)= (k-q)_{\alpha}g_{\mu\nu}
-(2k+q)_{\mu}g_{\nu\alpha} + (2q+k)_{\nu}g_{\mu\alpha}
\EQN Vertex$$
in the following way:
$$
\Gamma_{\mu\nu\alpha}=\Gamma^{P}_{\mu\nu\alpha} + \Gamma^{F}_{\mu\nu\alpha}
\EQN tHooft$$
with
$$\eqalign{
&\Gamma^{P}_{\mu\nu\alpha}=(q+k)_{\alpha}g_{\mu\nu}+
 k_{\nu}g_{\mu\alpha} ~~,\cr
&\Gamma^{F}_{\mu\nu\alpha}= 2q_{\nu}g_{\mu\alpha} -
 2q_{\alpha}g_{\mu\nu} -(2k+q)_{\mu}g_{\nu\alpha} ~.\cr}
\EQN Decomp$$
$\Gamma^{P}_{\mu\nu\alpha}$ (P for "pinch") gives rise to pinch
parts when contracted with $\gamma$ matrices in the fermion lines
\reference{WI}
$\Gamma^{F}_{\mu\nu\alpha}$ satisfies a Feynman-gauge Ward identity,
namely
$q^{\mu}\Gamma^{F}_{\mu\nu\alpha} = [k^2-(k+q)^2]g_{\nu\alpha}$,
where the r.h.s. is the difference of two inverse propagators.
\endreference
{}.
In the $\xi=1$ gauge, the relevant contributions from
$\Gamma^{P}_{\mu\nu\alpha}$ involve the factors
$$\eqalign{
&g_{\mu\nu}(\slashchar{q}+\slashchar{k})=
ig_{\mu\nu}[S^{-1}(p+q)-S^{-1}(p-k)] ~,\cr
&~~~~~~g_{\mu\alpha}\slashchar{k}=
ig_{\mu\alpha}[S^{-1}(p)-S^{-1}(p-k)] ~.\cr}
\EQN Pinch$$
Now both
$S^{-1}(p+q)$ and $S^{-1}(p)$ vanish on shell, whereas the two terms
proportional to $S^{-1}(p-k)$ "pinch" the internal fermion propagator
in graph 1b. In the $\xi=1$ gauge,
the total pinch contribution
from graph 1b and its mirror
image, with the vertex attached to the bottom fermion line, is
$-\Pi^{P}(q){\cal{M}}^{0}$, where ${\cal{M}}^{0}$ is the zeroth order
amplitude and
$$\eqalign{
\Pi^{P}(q)=&-(\frac{N}{2})\times 2 \times 2 \times
ig^2 {\mu}^{4-n}
\int\mathinner{[\frac{d^nk}{(2\pi)^n}]\frac{1}{k^2(k+q)^2}}\cr
=& \frac{2Ng^2}{16\pi^2}\ln(\frac{-q^2}{\mu^2}) .\cr}
\EQN TotalPinch$$
In the second equality we give the renormalized version of the
integral in the $\overline{MS}$ scheme
\reference{factors}
The factors in front of the integral are: a group-theoretic factor
 $\frac{1}{2}N$ [ $N$ = number of colors in $SU(N)$ ]; one factor of
2 for
the two pinching terms from \Eq{Pinch}; another factor
of 2 from the contribution of the mirror graph
\endreference
{}.
As this pinch part is proportional to ${\cal{M}}^{0}$ and only depends
on the momentum transfer $q^{2}$, it can be combined with the usual
self-energy contribution (Fig.1a). In fact, calling
$\Pi_{\mu\nu}(q)$ the gluon vacuum polarization tensor,
$-\Pi^{P}(q){\cal{M}}^{0}$ is equivalent to an additional contribution
of the form
$$
\Pi^{P}_{\mu\nu}(q)= P_{\mu\nu}(q)\Pi^{P}(q) ,
\EQN TransPinch$$
where
$$
P_{\mu\nu}(q) \equiv q_{\mu}q_{\nu}-q^{2}g_{\mu\nu}
\EQN ProjOper$$
is the transverse projection operator.
Following the conventions of
\reference{M&S}
W.~J.~Marciano and A.~Sirlin,
\journal Phys. Rev.; D 22,2695 (1980)
\endreference
and
\reference{D&SF}
G.~Degrassi and A.~Sirlin,
\journal Nucl. Phys.; B 383,73 (1992)
\endreference
,
in \Eq{TransPinch}
and henceforth we define the vacuum polarization tensor for a gauge boson
as ($-i$) times the sum of the one-particle irreducible self-energy diagrams.
Because we are working with the on-shell S-matrix and the vector currents
attached to the external fermions are conserved, any terms
$\sim q_{\mu}q_{\nu}$ in the pinch parts do not contribute to
${\hat{T}}_{1}(t)$. In \Eq{TransPinch} we have defined uniquely
the proper self-energy tensor associated with the
pinch parts by requiring that it be conserved
\reference{african shield}
This conserved form is, in fact, automatic in other forms of the pinch
technique, e.g., the off-shell approach of Ref.\cite{Kursunoglu}
or the intrinsic pinch discussed in
the fourth paper
of Ref.\cite{Bib}
\endreference
{}.
Adding \Eq{TransPinch}
to the conventional self-energy evaluated in the $\xi=1$
gauge,
$$
\Pi^{(\xi=1)}_{\mu\nu}(q)= \Pi^{(\xi=1)}(q) P_{\mu\nu}(q) ,
\EQN FeynGaugeTens$$
with
$$
\Pi^{(\xi=1)}(q)= (\frac{5N-2n_{f}}{3})\frac{g^2}{16\pi^2}
\ln(\frac{-q^2}{\mu^2}) ,
\EQN FeynGaugeScal$$
where $n_{f}$ is the number of effective quark flavors,
we find that the PT vacuum polarization
tensor is given by the gauge invariant expression:
$$
\hat{\Pi}_{\mu\nu}(q)= \hat{\Pi}(q)P_{\mu\nu}(q) ,
\EQN TransProp$$
with
$$
\hat{\Pi}(q)= \frac{bg^2}{16\pi^2}\ln(\frac{-q^2}{\mu^2}) ,
\EQN RunnCoupl$$
where
$b= \frac{11N-2n_{f}}{3}$ is
the coefficient of $- \frac{g^3}{16\pi^2}$
in the one loop $\beta$ function.
Including leading logarithms, the PT propagator reads
$$
{\hat{\cal{D}}}_{\mu\nu}(q)= \frac{i}{q^4}
\{\frac{P_{\mu\nu}(q)}
{[1+\frac{bg^2}{16\pi^2}\ln(\frac{-q^2}{\mu^2})]}-
 \xi q_{\mu}q_{\nu}\} .
\EQN FullProp$$
We see that the modified propagator has a gauge independent self-energy
that conforms with the renormalization group result,
and only a trivial gauge dependence originating from its tree
level part.

As a second application, we illustrate an
alternative formulation of the P.T. introduced in Ref.\cite{D&S}
in the context of the SM. In this
approach the interaction of gauge bosons with external fermions
is expressed in terms of
current correlation functions, i.e. matrix elements of Fourier transforms
of time-ordered products of current operators
\reference{Another Sirlin}
A.~Sirlin,
\journal Rev. Mod. Phys.; 50,573 (1978)
\endreference
{}.
This is particularly economical because these amplitudes automatically
include several closely related Feynman diagrams. When one of the current
operators is contracted with its four-momentum
(i.e. the four momentum absorbed by the current), a Ward identity
is triggered. The pinch part is then identified with the contributions
involving equal-time commutators in the Ward identities, and therefore
involve amplitudes in which the number of current operators has been
decreased by one or more.
As emphasized in Ref.\cite{D&S}, this procedure has an important advantage
when one considers external particles endowed with strong interactions.
Because the contributions from the equal-time commutators are not affected
by the dynamics of the strong interactions, the aforementioned
identification ensures the universality of the "pinch parts". That is, the
cofactors of the current operators in the pinch parts are the same whether
the external particles are leptons or strongly interacting fermions.
To illustrate the method with an example, consider
the vertex function $iU^{\lambda}_{z}(W)$ that  contributes to
Fig.(1b),
where now the gauge particles in the loop are W's, the incoming
one is a $Z$,
and the incoming and outgoing fermions are massless.
It can be written as
\reference{Error}
There is a misprint in Eq.(21) of Ref.\cite{D&SF} and Eq.(9a)
of Ref.\cite{D&S}, which does not affect the results of those papers:
an overall factor of $\frac{1}{\sqrt{2}}$ should be replaced by
$\frac{1}{2}$, and is corrected in Eq.(2.22) of the present paper
There is also a typographical error in Eq.(19) of the last paper:
a factor $q^{4}-m_{W}^{2}$ should be replaced by $q^{2}-m_{W}^{2}$
\endreference
$$\eqalign{
iU_{z}^{\lambda}(W)=&\frac{ig^{3}c}{2}
\int \frac{d^{n}k}{{(2\pi)}^n}
{\cal{D}}_{\alpha\rho}^{W}(k){\cal{D}}_{\sigma\beta}^{W}(k+q)
[g^{\rho\sigma}(2k+q)^{\lambda}-g^{\lambda\sigma}(2q+k)^{\rho}
-g^{\rho\lambda}(k-q)^{\sigma}]\cr
&\times\int d^{n}x e^{ikx}
<f|T^{*}[J^{\alpha\dagger}_{W}(x)J^{\beta}_{W}(0)]|i> ,\cr}
\EQN Papous$$
where henceforth $c$ and $s$ are abbreviations for $cos\theta_{w}$
and $sin\theta_{w}$, respectively.
{}~Choosing ~$\xi=1$ in the W propagators, the only longitudinal momenta
are provided by the trilinear gauge boson vertex.
An appropriate momentum, say $k_{\alpha}$,
from the vertex can be transformed into a derivative
 $\frac{d}{dx_{\alpha}}$ acting on the $T^{*}$ product.
Invoking current conservation
this leads to an equal-time commutator of current operators.
Thus, such contribution are proportional to the
matrix
element of a single current operator,
 namely $<f|J_{3}^{\lambda}|i>$; these
are precisely the pinch parts. Calling $i{U_{z}^{\lambda}(W)}_{P}$
 the total pinch contribution from
\Eq{Papous}, we find in the  $\xi=1$ gauge
$$
{U_{z}^{\lambda}(W)}_{P}= ig^{3}c
<f|J_{3}^{\lambda}|i>
{\mu}^{4-n}\int \frac{d^{n}k}{(2\pi)^{n}}
\frac{1}{(k^{2}-M_{w}^{2})[{(k+q)}^{2}-M_{w}^{2}]}.
\EQN PinchPapou$$
Clearly, the integral in \Eq{PinchPapou} is the generalization
of the QCD expression (\Eq{TotalPinch})
to the massive gauge boson case.

\section {Renormalizable $W$ self-energy in the unitary gauge.}
As is well-known,
in the unitary gauge ($\xi_{w} , \xi_{z} \rightarrow\infty$)
the $W$ and $Z$ propagators
are given by
$$
{\cal D}_{\mu\nu}^{i}(q)= [g_{\mu\nu} - \frac{q_{\mu}q_{\nu}}
{M^{2}_{i}}]\frac{-i}{q^{2}-M^{2}_{i}} ~,
\EQN UnitaryProp$$
where $i=W,Z$.
Because of the presence of the $q_{\mu}q_{\nu}$ term,
${\cal D}_{\mu\nu}^{i}(q)\sim 1 $ as $q\rightarrow\infty$ and,
as a consequence, the unitary gauge one-loop amplitudes are highly divergent.
If dimensional regularization is applied, this hard short distance behaviour
manifests itself in the occurrence of divergences proportional to high
powers of $q^{2}$. For example, keeping the photon propagator in the
Feynman gauge $(\xi_{\gamma}=1)$, but employing the unitary gauge for the
W and Z bosons, one obtains for the transverse W self-energy
\reference{Bardin}
D.~Yu.~Bardin, P.~Ch.~Christova and O.~M.~Fedorenco,
\journal Nucl. Phys.; B 175,435 (1980)
\endreference
$$
{\cal{A}}_{ww}(q^{2})_{div}
= \frac{g^{2}}{16\pi^{2}\epsilon}\{\frac{q^{6}c^{2}}{M^{2}_{w}M^{2}_{z}}
+q^{4}[\frac{5}{3}(\frac{1}{M^{2}_{w}}+\frac{c^{2}}{M^{2}_{z}})
-\frac{c^{2}}{2}(\frac{1}{M^{2}_{w}}+\frac{1}{M^{2}_{z}})]+...\}~,
\EQN Bardin$$
where $\epsilon=n-4$, the subscript "div" means "divergent part", and the
ellipses denote divergent contributions of $O(q^{2}, (q^{2})^{0})$.
In \Eq{Bardin} and henceforth, ${\cal{A}}_{ww}(q^{2})$ represents the
cofactor of $g_{\mu\nu}$ in the vacuum polarization tensor
$\Pi_{\mu\nu}^{ww}(q)$.
As we will see,\Eq{Bardin} can also be derived from
the results of Ref.\cite{D&SF}. We note that
 the divergent terms proportional
to $q^{6}$ and  $q^{4}$ in \Eq{Bardin} cannot be removed by the usual
mass and field-renormalization counter-terms. In this sense, the W-self energy
evaluated in the unitary gauge is "non-renormalizable".

We now turn our attention to the divergent pieces associated with the
charged current vertex and box diagrams in four-fermion processes.
(For
simplicity, we assume that the external fermion masses $m_{f}$ are very small
and take the limit $m_{f}\rightarrow 0$ whenever this procedure does not lead
to mass-singularities). In a renormalizable gauge, the vertex parts are
logarithmicaly divergent and, as a consequence, their divergent parts are
independent of $q^{2}$. Similarly, the box diagrams in four-fermion processes
are convergent. In the unitary gauge, the hard high-energy behaviour manifests
itself in the occurrence of divergent vertex parts proportional
to $q^{2}$ and  $q^{4}$, and to divergent box contributions proportional to
$(q^{2})^{0}$ and $q^{2}$. Such "non-renormalizable" divergent terms
can be readily obtained from Ref.\cite{D&SF} in the following way. In that
work, the self-energy, vertex and box diagrams of four-fermion processes
are expressed as the result in the 't-Hooft-Feynman gauge plus terms that
represent the additional contributions in an arbitrary $R_{\xi}$ gauge
characterized by the three parameters $\xi_{w}$, $\xi_{z}$, $\xi_{\gamma}$
\cite{D&SF}. The latter are given by integrals over virtual momenta, and
of course, they depend on $\xi_{w}$, $\xi_{z}$, $\xi_{\gamma}$. The results of
Ref.\cite{D&SF} are obtained after neglecting longitudinal terms
proportional to the momentum transfer $q^{\lambda}$ which give vanishing
contributions to the S-matrix in the $m_{f}\rightarrow 0$ limit.
Otherwise, their derivation involves only the application of Ward
identities and algebraic manipulations, without carrying out explicit
integrations over the loop momenta.
In the case of vertex and box diagrams,
the additional terms mentioned above are
pinch parts, as they originate from amplitudes involving one or more
equal-time commutators of current operators. We recall that such
contributions are triggered by associated Ward identities.
In order to obtain the desired results in the unitary gauge, we take the
limit $\xi_{w} \rightarrow\infty$, $\xi_{z} \rightarrow\infty$ in the
pinch parts (keeping $\xi_{\gamma}=1$),
{\sl before} performing the integration over
virtual momenta. After this is done, the resulting  integrals are of the
form
$$
it^{\mu\nu}\mu^{4-n}
\int\frac{d^{n}k}{(2\pi)^{n}}\frac{[g_{\mu\nu};k_{\mu}k_{\nu}]}
{(k^{2}-M_{w}^{2})[{(k+q)}^{2}-M_{z}^{2}]}=
\frac{1}{16\pi^{2}\epsilon}
[2;\frac{M^{2}_{w}+M^{2}_{z}}{2}-\frac{q^{2}}{6}]+... ~,
\EQN TwoCases$$
where $t^{\mu\nu}=\frac{1}{n-1}
(g^{\mu\nu}-\frac{q^{\mu}q^{\nu}}{q^{2}})$
projects the $g_{\mu\nu}$ components
of the integrals, and the ellipses denote finite terms.
Using Eqs (33, 36, 37)
and the definitions in Eqs(2-5) of Ref.\cite{D&SF}, as well as
\Eq{TwoCases}
of the present paper, we readily find that the "non-renormalizable" divergent
vertex part is given by
$$
i\Gamma^{\lambda}_{w}|_{div}=
\frac{ig^{3}}{2\sqrt{2}}
\frac{<f|J_{w}^{\lambda}|i>}{16\pi^{2}\epsilon}
\{\frac{q^{4}c^{2}}{3M^{2}_{w}M^{2}_{z}}
+q^{2}[\frac{11}{3}
(\frac{1}{M^{2}_{w}}+\frac{c^{2}}{M^{2}_{z}})
-c^{2}(\frac{1}{M^{2}_{w}}+\frac{1}{M^{2}_{z}})]+...\}~,
\EQN VertexDiv$$
where the ellipses represent pieces involving lower powers of $q^{2}$.
Similarly, employing Eqs.(36, 37, 2-5) of Ref.\cite{D&SF}, we find that
the "non-renormalizable" divergent part of the box diagrams is of the
form
$$\eqalign{
B^{(CC)}|_{div}&=\frac{-ig^{4}}{32\pi^{2}\epsilon}
<f|J_{w}^{\lambda}|i><f'|{J_{w}^{\dagger}}_{\lambda}|i'>\cr
&\times \{-\frac{q^{2}c^{2}}{6M^{2}_{w}M^{2}_{z}}
-2(\frac{1}{M^{2}_{w}}+\frac{c^{2}}{M^{2}_{z}})
+\frac{c^{2}}{2}(\frac{1}{M^{2}_{w}}+\frac{1}{M^{2}_{z}})+...\}~.\cr}
\EQN BoxDiv$$
Recalling Eqs.(10,2-5) of Ref.\cite{D&SF}, one can apply the same method to
verify \Eq{Bardin}. As \Eq{VertexDiv} and \Eq{BoxDiv} are pinch parts,
in the PT their contributions are removed from the vertex and box diagrams
and combined with the usual self-energy. Taking into account their respective
contributions to the S-matrix, one finds that the proper contribution is
$$
{\hat{\cal{A}}}_{ww}(q^{2})_{div}={\cal{A}}_{ww}(q^{2})_{div}
-\frac{g^{2}}{16\pi^{2}\epsilon}(q^{2}-M^{2}_{w})\{V\}
-\frac{g^{2}}{16\pi^{2}\epsilon}{(q^{2}-M^{2}_{w})}^{2}\{B\}~,
\EQN PinchDiv$$
where $\{V\}$ and $\{B\}$ denote the expressions between curly brackets in
\Eq{VertexDiv} and \Eq{BoxDiv}, respectively.
Using \Eq{Bardin}, \Eq{VertexDiv}, and \Eq{BoxDiv} one readily verifies
that all the divergent terms involving $q^{4}$ and $q^{6}$
automatically cancel in ${\hat{\cal{A}}}_{ww}(q^{2})_{div}$.
Thus, the PT self-energy ${\hat{\cal{A}}}_{ww}(q^{2})_{div}$ is
"renormalizable" in the sense that its divergent parts can be removed by the
usual mass and field-renormalization counter-terms.

\section{PT W Self-Energy in the Unitary Gauge.}

In the previous section we showed that application of the PT algorithm in
unitary gauge calculations leads to a renormalizable modified W self-energy.
In this section we explicitly demonstrate that the renormalizable W
self-energy so constructed is {\sl identical} to the one obtained in
Ref.\cite{D&S} in the context of the linear, renormalizable $R_{\xi}$ gauges.
We consider here the complete one-loop expressions, involving both divergent
and finite parts, and furthermore we include the tadpole contributions in the
definition of the self-energy ${\cal{A}}_{ww}(q^{2})$. Our strategy is
analogous to that employed in Section 3. Using Eqs.(10) and (12) of
Ref.\cite{D&SF}, and taking the limit
$\xi_{w} , \xi_{z} \rightarrow\infty$ (with $\xi_{\gamma}=1$ fixed)
in $[{\cal{A}}_{ww}(q^{2})-{\cal{A}}_{ww}(q^{2})|_{\xi_{i}=1}]$,
{\sl without} performing the integration over virtual momenta,
we obtain
$$
{\cal{A}}_{ww}(q^{2})={\cal{A}}_{ww}(q^{2})|_{\xi_{i}=1}+
g^{2}(q^{2}-M^{2}_{w})L_{1}(q^{2})+
 g^{2}{(q^{2}-M^{2}_{w})}^{2}L_{2}(q^{2})~,
\EQN DS1$$
where
$$\eqalign{
L_{1}(q^{2})=it^{\mu\nu}&\int_{n}\Biggl\lbrack
-\frac{1}{M^{2}_{w}(k^{2}-M^{2}_{w})}+
\frac{c^{2}}{M^{2}_{w}}\{2g_{\mu\nu}(q^{2}-M^{2}_{z})+
\frac{k_{\mu}k_{\nu}}{M^{2}_{z}}(2M^{2}_{z}-q^{2})\}\Sigma_{wz}\cr
&+ (W \leftrightarrow Z) + 2\frac{s^{2}}{M^{2}_{w}}
[k_{\mu}k_{\nu}+ q^{2}g_{\mu\nu}]\Sigma_{w\gamma}
\Biggr\rbrack ~,\cr}
\EQN L1$$
$$
L_{2}(q^{2})=\frac{1}{2}it^{\mu\nu}\int_{n}
[\frac{c^{2}}{M^{2}_{w}}(\frac{k_{\mu}k_{\nu}}{M^{2}_{z}}-2g_{\mu\nu})
\Sigma_{wz}
+ (W \leftrightarrow Z)-\frac{2s^{2}}{M^{2}_{w}}g_{\mu\nu}
\Sigma_{w\gamma}]~,
\EQN L2$$
where
$$
\int_{n}\equiv {\mu}^{4-n}\int \frac{d^{n}k}{(2\pi)^{n}} ~,
\EQN In$$
and
$$
\Sigma_{ij}\equiv \frac{1}{(k^{2}-M_{i}^{2})[{(k+q)}^{2}-M_{j}^{2}]}
{}~~~(i,j=W,Z,\gamma)~.
\EQN BasicRatiol$$
In \Eq{L1} and \Eq{L2} the notation $(W \leftrightarrow Z)$
represents terms obtained from the previous ones by the substitution
$M_{w}\leftrightarrow M_{z}$ and, in \Eq{BasicRatiol},
 $M_{\gamma}=0$.
Applying the same procedure to Eqs.(33), (36), and (37) of
Ref.\cite{D&SF}, we have for the vertex and box diagrams:
$$
i\Gamma_{w}^{\lambda}=i\Gamma_{w}^{\lambda}|_{\xi=1}
+i\frac{g^{3}}{2\sqrt{2}}
<f|J_{w}^{\lambda}|i>L_{1}(q^2)~,
\EQN FullGamma$$
$$
B^{CC}=B^{CC}|_{\xi=1}-\frac{ig^{4}}{2}L_{2}(q^2)
<f|J_{w}^{\lambda}|i><f'|{J_{w}^{\dagger}}_{\lambda}|i'>~,
\EQN FullBox$$
where $L_{1}(q^2)$ and $L_{2}(q^2)$ are the same functions given
in \Eq{L1} and \Eq{L2}. We recall that, because of the momenta present in
the trilinear gauge boson vertex, $\Gamma_{w}^{\lambda}|_{\xi=1}$ contains
pinch parts \cite{D&S}.
Specifically:
$$\Gamma_{w}^{\lambda}|_{\xi=1}=\Gamma_{w}^{\lambda}|_{NP}+
g^{3}\sqrt{2}[c^{2}I_{zw}(q^2)+s^{2}I_{\gamma w}(q^2)]
<f|J_{w}^{\lambda}|i>~,
\EQN GammaFeyn$$
where $\Gamma_{w}^{\lambda}|_{NP}$ stands for the non-pinch part, and
$$
I_{ij}(q^2)= \int_{n}\Sigma_{ij}~~.
\EQN DefI$$
Inserting \Eq{GammaFeyn} into \Eq{FullGamma}, we separate the complete
pinch part of $\Gamma_{w}^{\lambda}$:
$$
i\Gamma_{w}^{\lambda}=i\Gamma_{w}^{\lambda}|_{NP}+
\frac{ig^{3}}{2\sqrt{2}}
[L_{1}(q^2)+4(c^{2}I_{zw}(q^2)+s^{2}I_{\gamma w}(q^2)]
<f|J_{w}^{\lambda}|i>~.
\EQN FinalGamma$$
On the other hand, there are no pinch parts in $B^{CC}|_{\xi=1}$ so that
$B^{CC}|_{\xi=1}=B^{CC}|_{NP}$. According to the PT, the pinch part of
$i\Gamma_{w}^{\lambda}$ ( the second term in \Eq{FinalGamma})
 and the pinch
part of  $B^{CC}$ (the second term in \Eq{FullBox}) are removed from
$i\Gamma_{w}^{\lambda}$ and $B^{CC}$, respectively, and combined with
${\cal{A}}_{ww}(q^{2})$.The remaining contributions,
$i\Gamma_{w}^{\lambda}|_{NP}$ and $B^{CC}|_{NP}$, are identified with the
PT vertex and box diagrams, respectively. Taking into account their respective
contributions to the S-matrix, one finds that the appropriate combination of
the self-energy with the pinch parts derived from $i\Gamma_{w}^{\lambda}$
and $B^{CC}$ is
$$
{\hat{\cal{A}}}_{ww}(q^{2})={\cal{A}}_{ww}(q^{2})-
g^{2}(q^{2}-M^{2}_{w})[L_{1}(q^2)+4\{c^{2}I_{zw}(q^2)+
s^{2}I_{\gamma w}(q^2)\}]
-g^{2}{(q^{2}-M^{2}_{w})}^{2}L_{2}(q^2)
\EQN AlmostEnd$$
Recalling \Eq{DS1}, we see that the terms involving $L_{1}(q^2)$ and
$L_{2}(q^2)$ exactly cancel and we finally obtain
$$
{\hat{\cal{A}}}_{ww}(q^{2})= {\cal{A}}_{ww}(q^{2})|_{\xi=1}
-4g^{2}(q^{2}-M^{2}_{w})[c^{2}I_{zw}(q^2)+s^{2}I_{\gamma w}(q^2)]~.
\EQN End$$
Comparing \Eq{End} with Eq.(19) of Ref.\cite{D&SF}, we see that, in fact,
the self-energy ${\hat{\cal{A}}}_{ww}(q^{2})$ obtained by applying the PT
to the unitary gauge calculation is {\sl exactly} the same as the expression
previously derived in Ref.\cite{D&S} in the context of the $R_{\xi}$ gauges.
The same holds true for the non-pinch parts
$\Gamma_{w}^{\lambda}|_{NP}$ and $B^{CC}|_{NP}$. It is worth noting that the
proof in this section involves only the application of Ward identities and
algebraic manipulations, without carrying out explicit integrations over
loop momenta.

\section{References}
\ListReferences
\section{Figure Caption}
Graphs (a)-(c) are some of the contributions to the S-matrix
for four-fermion processes.
Graphs (e) and (f) are pinch parts, which, when added to the usual
self-energy graphs (d), give rise to the gauge-independent amplitude
${\hat{T}}_{1}(t)$.
The mirror image graph corresponding to (b) is not shown.

\bye